\documentstyle[epsf]{elsart}

\def\0{\over } \def\2{\textstyle{1\over2}} \def\4{\textstyle{1\over4}}
\def\5{\hat } \def\6{\partial }

\let\a=\alpha \let\b=\beta  

 \let\o=\omega
   
\let\O=\Omega

\def\({\left(} \def\){\right)} \def\<{\langle } \def\>{\rangle }
\def\[{\left[} \def\]{\right]}  
\newcommand{\bea}{\begin{eqnarray}}
\newcommand{\eea}{\end{eqnarray}}
\newcommand{\be}{\begin{equation}}
\newcommand{\ee}{\end{equation}}
\newcommand{\nn}{\nonumber\\ }
\newcommand \beq{\begin{eqnarray}}
\newcommand \eeq{\end{eqnarray}}

\def\simge{\mathrel{%
   \rlap{\raise 0.511ex \hbox{$>$}}{\lower 0.511ex \hbox{$\sim$}}}}
\def\simle{\mathrel{
   \rlap{\raise 0.511ex \hbox{$<$}}{\lower 0.511ex \hbox{$\sim$}}}}

\def\Tr{{\,\mathrm Tr\,}}
\def\Im{{\,\mathrm Im\,}}
\def\Re{{\,\mathrm Re\,}}
\def\tr{{\,\mathrm tr\,}}

\begin{document}
\begin{flushright}
~\vspace{-1.25cm}\\
{\small\sf
SACLAY-T99/113\\ CERN-TH/99-307\\ TUW-99/21\\
}
\end{flushright}
\vspace{0.2cm}
\begin{frontmatter}
\title{Self-consistent hard-thermal-loop thermodynamics
for the quark-gluon plasma\thanksref{Amadeus}}
\thanks[Amadeus]{Work supported by the Austrian-French
scientific exchange program Amadeus of APAPE and \"OAD}

\author[Saclay]{J.-P. Blaizot}, \author[CERN]{E. Iancu} and
\author[Wien]{A. Rebhan}
\address[Saclay]{Service de Physique Th\'eorique, CE Saclay,
        F-91191 Gif-sur-Yvette, France}
\address[CERN]{Theory Division, CERN, CH-1211 Geneva 23, Switzerland}
\address[Wien]{Institut f\"ur Theoretische Physik,
         Technische Universit\"at Wien,\\
         Wiedner Hauptstra\ss e 8-10/136,
         A-1040 Vienna, Austria}

\begin{abstract}
Self-consistent approximations allowing the calculation of the entropy 
and the baryon density of a quark-gluon plasma
are presented. These approximations incorporate the essential physics of
the hard thermal loops, involve only ultraviolet-finite  quantities, 
and are free from overcounting ambiguities. 
While being nonperturbative in the strong coupling constant $g$,
agreement with ordinary perturbation theory is achieved
up to and including order $g^3$. It is shown how 
the pressure can be reconstructed from the entropy and the baryon density
taking into account the scale anomaly.
The results obtained are in good agreement with available
lattice data down to temperatures
of about twice the critical temperature.

\end{abstract}
\end{frontmatter}

\section{Introduction}

While asymptotic freedom suggests that, at very high temperature and/or
sufficiently large baryon  
density, QCD should behave as a weakly coupled quark-gluon plasma, 
lattice results reveal that this ideal gas limit is approached only 
rather slowly.
Perturbation theory \cite{Kapusta,QCDP}
fails to reproduce this behaviour correctly, 
and resummations beyond the conventional ring-resummation
are needed to properly account for 
the thermodynamics. In QCD,
the situation may not be hopelessly complicated however; in fact, the
available lattice data are quite
well reproduced by simple models based on 
non-interacting ``quasiparticles''  with
temperature dependent masses \cite{Peshier,LH}, 
which suggests that it could be
possible to describe the thermodynamics of the quark-gluon plasma 
in terms of its elementary
excitations. 

However, phenomenological fits using massive quasiparticles represent  only a
convenient parametrisation of the thermodynamics and further theoretical effort is needed  to
arrive at a complete understanding 
in terms of the elementary degrees of freedom and their
interactions.  What the quasiparticle picture suggests is 
that much of the physics at high
temperature can presumably be described by  propagator
renormalisation  or, in other words, 
a  {\it resummed} perturbation theory in terms of {\it
dressed} propagators. The importance of retaining their effects
nonperturbatively has been demonstrated recently in
scalar toy models \cite{KPP,DHLR2}. The general framework for 
carrying out consistently such renormalisations
 of the thermodynamical functions has been well 
developed in the past in the context of the 
non-relativistic many-body problem 
\cite{LW,Baym,CP2}.  

A convenient quantity to consider when doing 
propagator renormalisation is the {\it entropy}, 
rather than the pressure \cite{CP2,Riedel}. Entropy is better
suited, indeed, because it has a  more direct  
interpretation in terms of quasiparticles; in
particular, it is  expressible entirely in 
terms of dressed propagators, at least in lowest
orders in a skeleton expansion. Building upon 
such an expansion, we are able to
construct approximations which incorporate,
in a way that is free of overcounting ambiguities,
all the important physics of the dominant
contributions to the self-energies at high temperature, the  so-called ``hard
thermal loops'' (HTL) \cite{BP,MLB}. (Corresponding vertex corrections, 
a priori needed to maintain
gauge symmetry, turn out to be unimportant at the order of interest.) 
In contrast to 
the direct 
HTL resummation of the pressure recently proposed in
Ref.~\cite{ABS}, the resummation that
we perform involves only UV-finite expressions and when
compared to  perturbation theory it agrees up to, and including, order
$g^3$. 

In Ref.~\cite{PRL}, this strategy has been 
successfully applied to estimate the entropy ${\cal S}$ of
the purely Yang-Mills plasma. 
In this Letter, we consider two important extensions. The first
one  is the inclusion of quarks which allows us to also
treat plasmas with a
finite  chemical potential $\mu$. 
The relevant formalism, presented in Secs. 2 and 3
below, leads to  self-consistent approximations for both the
entropy ${\cal S}$ and the baryon density ${\cal N}$.
The second extension concerns the computation of thermodynamical
quantities other than the entropy or the density. In Sec. 4, we show
how to reconstruct the pressure ${P}$ by a numerical integration
of  ${\cal S}$ and ${\cal N}$;  because
of the trace anomaly, this requires some nonperturbative 
input which we take from
lattice simulations at $\mu=0$. 
For all temperatures $T\simge 2\ldots3T_c$, our results 
are in remarkable agreement with the corresponding lattice results whenever
the latter are available. 
We also present results for the baryon density at $T=0$ and large $\mu$.
Further possible extensions of our method 
are outlined in Sec. 5.

\section{Self-consistent approximation to the entropy and quark density}

The thermodynamic potential $\Omega=-PV$ of a field theory involving
bosons and fermions, expressed as a functional of full propagators
($D$ for bosons, $S$ for fermions) has the form \cite{LW}
\bea\label{LWQCD}
\!\!\!\!\!\!\!\!
\O[D,S]&=&\2 T \Tr \log D^{-1}-\2 T \Tr \Pi D
- T \Tr \log S^{-1} + T\Tr \Sigma S + T\Phi[D,S] \nn
&=& T\Phi[D,S] + \tr \int{d^4k\0(2\pi)^4}n(\omega) \Im \left[
\log D^{-1}(\omega,k)-\Pi(\omega,k) D(\omega,k) \right] \nn
&&+2\tr \int{d^4k\0(2\pi)^4}f(\omega) \Im \left[
\log S^{-1}(\omega,k)-\Sigma(\omega,k) S(\omega,k) \right]
\eea
where 
$\Phi[D,S]$ is the sum of 2-particle-irreducible ``skeleton''
diagrams, $n(\omega)=(\e^{\b\omega}-1)^{-1}$,
$f(\omega)=(\e^{\b(\omega-\mu)}+1)^{-1}$, and
$\beta=1/T$. Here ``tr'' refers to all discrete labels,
including colour and flavour when applicable.

The self-energies $\Pi=D^{-1}-D^{-1}_0$ and $\Sigma=S^{-1}-S^{-1}_0$,
where $D_0$ and $S_0$ are bare propagators, are themselves functionals
of the full propagators and are determined by
\be\label{selfenergies}
{\delta\Phi[D,S]/\delta D}=\half\Pi,\quad
{\delta\Phi[D,S]/\delta S}=\Sigma,
\ee
which goes hand in hand with the all-important stationarity property
\be\label{staty}
{\delta \Omega[D,S]/\delta D}=0={\delta \Omega[D,S]/\delta S}.
\ee

A ``self-consistent'' (``$\Phi$-derivable'') \cite{Baym} approximation
is one that preserves this stationarity property by selecting a subset
of skeleton-diagrams from $\Phi$ and determining the self-energies
from (\ref{selfenergies}).
In particular, a two-loop approximation to $\Phi[D,S]$
(corresponding to a dressed one-loop approximation for the
self-energies) is obtained by discarding all skeleton diagrams of
loop-order 3 and higher.

This two-loop approximation for $\Omega$ has a remarkable consequence
for the first derivatives of the thermodynamic potential, the entropy
and the fermion densities:
\be\label{SNdef}
{\cal S}=-{\6(\Omega/V)\0\6T}\Big|_{\mu},\quad
{\cal N}=-{\6(\Omega/V)\0\6\mu}\Big|_{T}.
\ee
Because of the stationarity property (\ref{staty}), one can ignore the $T$ and
$\mu$ dependences implicit in the spectral densities of the full
propagators, and differentiate exclusively the statistical
distribution functions $n$ and $f$ in (\ref{LWQCD}).
Now the derivative of the {\em two-loop} functional $T\Phi[D,S]$ at fixed
spectral densities of the propagators $D$ and $S$ 
turns out to just cancel
that part of the terms $\Im(\Pi D)$ and $\Im(\Sigma S)$ in
(\ref{LWQCD}) which
involves $\Re\Pi\Im D$ and $\Re\Sigma\Im S$, respectively.
In a {\em self-consistent} two-loop approximation one therefore has
the remarkably simple formulae
\bea
\label{S2loop}
{\cal S}&=&-\tr \int{d^4k\0(2\pi)^4}{\6n(\omega)\0\6T} \left[ \Im 
\log D^{-1}(\omega,k)-\Im \Pi(\omega,k) \Re D(\omega,k) \right] \nn
&&-2\tr \int{d^4k\0(2\pi)^4}{\6f(\omega)\0\6T} \left[ \Im
\log S^{-1}(\omega,k)-\Im \Sigma(\omega,k) \Re S(\omega,k) \right], \\
\label{N2loop}
{\cal N}&=&-2\tr \int{d^4k\0(2\pi)^4}{\6f(\omega)\0\6\mu} \left[ \Im
\log S^{-1}(\omega,k)-\Im \Sigma(\omega,k) \Re S(\omega,k) \right].
\eea
This has been noted first for the entropy in a non-relativistic
context by Riedel \cite{Riedel} and more recently in QED by
Vanderheyden and Baym \cite{VB}, but 
this important simplification holds much more generally
and is equally valid for the fermion density \cite{us}.

In a nonabelian gauge theory, the above relations have to be in general
augmented by ghost contributions, which can however be
avoided in those gauges where ghosts do not propagate, such
as Coulomb gauge, in which we shall work in what follows.
At any rate, in gauge theories, abelian as well as nonabelian,
self-consistency does not guarantee
gauge invariance, for only
propagators have been dressed, and no vertices\footnote{Vertices
can be dressed in a self-consistent manner in the formalism
worked out in Ref.~\cite{FM}.}.

However, in what follows we shall consider {\em approximately
self-consistent} approximations which guarantee gauge independence.
These amount to compute the self-energies in (resummed) perturbation theory,
satisfying the self-consistency condition
in a perturbative sense. On the other hand, we shall not expand
out the self-energies from eqs. (\ref{S2loop})
and (\ref{N2loop}) so that the resulting approximations for the
${\cal S}$ and ${\cal N}$ remain {\em nonperturbative} in $g$.
This is possible without encountering ultraviolet divergences
because (in contrast to the original thermodynamic potential)
the expressions (\ref{S2loop}) and (\ref{N2loop}) have the
decisive advantage of being manifestly {\em ultraviolet
finite} since the derivatives of the distribution functions
vanish exponentially for both $\omega\to\pm\infty$. Moreover,
any multiplicative renormalization with real $Z$ leaves
(\ref{S2loop}) and (\ref{N2loop}) unchanged.

\section{Contact with perturbation theory}

In Coulomb gauge, the gluon
propagator consists of a transverse and a longitudinal
piece, with LO self-energies at soft momenta ($\omega,k\ll
{\rm max}(T,\mu)$) given by the so-called
hard thermal (dense) loops \cite{BP,MLB}
\bea
\label{PiTL}
\5\Pi_L(\o,k)&=&\5m_D^2\left[1-{\o\02k}\log{\o+k\0\o-k}\right],\nn
\5\Pi_T(\o,k)&=&\frac{1}{2}\left[\5m_D^2+\,\frac{\o^2 - k^2}{k^2}\,
\5\Pi_L\right],
\eea
where
\be\label{MD}
\hat m^2_D=(2N+N_f)\,\frac{g^2T^2}{6}\,+\,N_f\,{g^2\mu^2\02\pi^2}
\ee
is the (leading-order) Debye screening mass.\footnote{For simplicity
we shall write out our formulae for only one value of the chemical
potential; the generalization to several different chemical potentials is
straightforward. }
As is well known \cite{KKW}, the gluon propagator dressed by these self-energies
has quasiparticle poles at $\o_{T,L}(k)$
with momentum-dependent effective masses
and Landau damping cuts for $|\omega|<k$. When $k\gg \5m_D$, the
pole corresponding to the collective longitudinal
excitation has exponentially vanishing residue \cite{P}, whereas 
that of the
transverse excitations tend to $\sqrt{k^2+m_\infty^2}$, where 
the asymptotic mass is given by
\be\label{mas}
m_\infty^2=\5\Pi_T(k,k)=\2 \5m_D^2\,.
\ee

The (massless) quark propagator at finite temperature or density is split into
two separate branches of opposite ratio of chirality over helicity
with propagators
$
\Delta_\pm=[-\o+k \pm \Sigma_\pm]^{-1}
$.
In the HTL approximation, the respective self-energies read
\cite{KKW,MLB}:
\be\label{SIGHTL}
\hat\Sigma_\pm(\omega,k)\,=\,{\5M^2\0k}\,\left(1\,-\,
\frac{\omega\mp k}{2k}\,\log\,\frac{\omega + k}{\omega - k}
\right),\ee
where ($C_f=(N^2-1)/2N$)
\be\label{MF}
\5M^2\,=\,{g^2 C_f\08}\left(T^2+{\mu^2\0\pi^2}\right).\ee

The dressed fermion propagators have quasiparticle poles 
at $\o_\pm(k)$ and Landau damping cuts.
At large $k$ and positive frequency 
$\o_+\to\sqrt{k^2+M_\infty^2}$
with asymptotic mass $M_\infty^2=2\5M^2$;
the abnormal branch $\o_-$ has exponentially vanishing residue \cite{P}. 

Although the various HTL self-energies constitute a leading-order
result only under the condition of {\em soft} momenta and frequencies,
the result for the asymptotic masses applies also for {\em hard} momenta
\cite{KKR}.

The LO interaction contribution $\propto g^2$ in our expressions for
the entropy and the density arise from the domain of hard momenta.
Because of the subtraction terms in (\ref{S2loop}) and (\ref{N2loop}),
it turns out \cite{PRL,us}
that this contribution is given entirely
in terms of the LO values for
$\Re\Pi_T(\o^2=k^2)$ and $\Re\Sigma_\pm(\o=\pm k)$,
yielding
\be\label{SN2}
{\cal S}^{(2)}=-N_g{m_\infty^2 T\06}-NN_f{M_\infty^2T\06},\quad
{\cal N}^{(2)}=-NN_f {M_\infty^2\,\mu\02\pi^2}
\ee
($N_g=N^2-1$)
in agreement with known results \cite{Kapusta}.\footnote{%
By contrast, the direct HTL resummation of the {\em pressure}
overincludes the LO interaction contributions $\propto g^2$
\cite{ABS,BR}.}
Notice the
simplicity of these results in comparison with conventional two-loop
calculations. Quite generally the LO interaction
results for entropy and fermion density are given by
${\cal S}^{(2)} =-  T\{{1\0 12} \sum_B m_{\infty\,B}^2 +
{1\0 24} \sum_F M_{\infty\,F}^2\},\quad
{\cal N}^{(2)}=-{1\0 8\pi^2}\sum_F \mu_F M_{\infty\,F}^2$
where the sums run over all bosonic (B) and fermionic (F) degrees of
freedom.
Thus the LO interaction terms in ${\cal S}$ and ${\cal N}$ are
entirely given by the quasiparticle dispersion relations at large 
momenta.
In fact, these formulae for ${\cal S}^{(2)}$ and ${\cal N}^{(2)}$
are identical to those corresponding to massive non-interacting
particles (with constant masses $m=m_\infty$ and $M_\infty$, respectively)
when expanded to leading order in $m^2$.

When $T\gg \5m_D$, the NLO contributions are of order
$g^3$ and in the conventional perturbation theory for the pressure arise
from the sum of (electro-static) ring diagrams,\footnote{If
$T\simle g\mu$, the
ring diagrams give rise to contributions of order $g^4\log(g)$.}
which yields
$P^{(3)}={1\012\pi}N_g \5m_D^3 T\propto g^3$.

If the propagators and self-energies in (\ref{S2loop}) and (\ref{N2loop})
are those of the HTL approximation, the order $g^3$ contributions
in the corresponding expressions for ${\cal S}$ and ${\cal N}$, denoted
${\cal S}_{\rm HTL}$ and ${\cal N}_{\rm HTL}$, turn out to
give only the part of the full result that is obtained by differentiating
$P^{(3)}$ at fixed $\hat m_D$ \cite{PRL,us}, ${\cal N}^{(3)}_{\rm HTL}=0$
and ${\cal S}^{(3)}_{\rm HTL}={1\012\pi}N_g \5m_D^3$.
The remaining contributions of order $g^3$ come from NLO corrections to the
asymptotic quasiparticle mass and are calculable in HTL perturbation
theory \cite{us}. 
These corrections are, in contrast to the LO asymptotic masses,
momentum-dependent, and increasing at hard momenta. In order to
estimate their contribution in our numerical evaluations, we shall
approximate them by averaged, effective NLO corrections
to $m_\infty$ and $M_\infty$ which are uniquely determined through
eq.~(\ref{SN2}) as
\cite{us}
\be\label{deltamas}
\bar\delta m_\infty^2=-{1\02\pi}g^2NT\hat m_D,\quad
\bar\delta M_\infty^2=-{1\02\pi}g^2C_fT\hat m_D.
\ee
In these expressions, the linear $\5m_D$ dependence and the
Casimir factors are indeed those expected from the corresponding
HTL-resummed one-loop diagrams.

\section{Nonperturbative evaluation --- recovering the pressure}

In our nonperturbative, numerical evaluation of (\ref{S2loop}) and
(\ref{N2loop}) we consider two successive approximations. The first
consists of using HTL-resummed propagators, which as we have discussed
reproduces the LO interaction terms and part of the NLO one.
As a second step we take into account the NLO corrections to the
asymptotic mass in their averaged form (\ref{deltamas}) and we do
so by $m_\infty^2\to \bar m_\infty^2 =
 m_\infty^2 [1-\bar\delta m_\infty^2/m_\infty^2]^{-1}$
and similarly for the fermionic asymptotic mass $M_\infty$.
This simple Pad\'e resummation can be shown to give a surprisingly
good approximation to the gap equation in exactly solvable scalar
models \cite{us}. However, because (\ref{deltamas}) refers to
the asymptotic mass and because the plasma frequency and the Debye
mass turn out to have rather different NLO corrections \cite{Sch,RDeb,LP},
we restrict this modification to hard momenta defined by
$k>\Lambda=\sqrt{2\pi T\5m_D c_\Lambda}$ and vary $c_\Lambda$ to
test the stability of the final results.

${\cal S}_{\rm HTL}$ can be separated in two physically distinct
contributions: one from quasiparticle poles, the other from Landau
damping cuts of the various excitations. The gluonic ones, which
only contribute to the entropy, have been
given in eqs. (16) and (17) in  Ref.~\cite{PRL}.  
The fermionic ones are quite analogous and read ${\cal S}_{f\,{\rm HTL}}
= {\cal S}_{f\,{\rm HTL}}^{\rm QP}+{\cal S}_{f\,{\rm HTL}}^{\rm LD}$
with
\bea
{\cal S}_{f\,{\rm HTL}}^{\rm QP}&=&2NN_f \int{d^3k\0(2\pi)^3}
{\6\0\6T} \Bigl\{ T\log(1+\e^{-[\o_+(k)-\mu]/T}) \nn
&&\qquad\qquad\qquad+ T\log{1+\e^{-[\o_-(k)-\mu]/T} \0 1+\e^{-(k-\mu)/T}} 
+(\mu\to-\mu) \Bigr\}
\eea
where only the explicit $T$ dependences are differentiated and not those
implicit in the dispersion laws $\o_+(k)$ and $\o_-(k)$ of the fermionic
quasiparticles.

The contribution to the quark density is obtained by replacing $\6/\6T$
in the above formula by $\6/\6\mu$. In the limit $T\to0$, the resulting
expression can be simplified to read ($\mu>0$)
\be
{\cal N}_{f\,{\rm HTL}}^{\rm QP}\Big|_{T=0} = NN_f 
\int_0^{\mu}
{k^2 dk\0\pi^2} [  \theta(\mu-\o_+(k)) - \theta(\o_-(k)-\mu) ].
\ee

The fermionic Landau-damping contribution to the entropy is
\bea
&&{\cal S}_{f\,{\rm HTL}}^{\rm LD}=-4NN_f \int{d\o\,d^3k\0(2\pi)^4}
{\6f(\o)\0\6T} \theta(k^2-\o^2) \Bigl\{ \arg[k-\o+\Sigma_+(\o,k)] \nn
&&\qquad\qquad\qquad\qquad\qquad\quad
-\Im\Sigma_+(\o,k) \Re[k-\o+\Sigma_+(\o,k)]^{-1} 
\nn&&\quad+\,
\arg[k+\o+\Sigma_-(\o,k)]-\Im\Sigma_-(\o,k) \Re[k+\o+\Sigma_-(\o,k)]^{-1}
\Bigr\}
\eea
and that to the quark density,
${\cal N}_{f\,{\rm HTL}}^{\rm LD}$, is again obtained by replacing
$\6/\6T$ by $\6/\6\mu$. In the limit of zero temperature we simply
have $\6f/\6\mu\to\delta(\o-\mu)$.

The coupling constant in the HTL masses $\5m_D$ and $\5M$ is
determined from the two-loop renormalization group equation
and as a central value for the ($\overline{\rm MS}$)
renormalization scale $\bar\mu$ we choose for the case of zero
chemical potential the spacing of the Matsubara frequencies $2\pi T$,
and for the case of zero temperature the diameter of the
Fermi sphere, $2\mu$. The QCD scale $\Lambda_{\overline{\rm MS}}$
is, according to lattice data, rather close to $T_c$
at $\mu=0$; for definiteness we adopt\footnote{This value is about
10 \% higher then the one used in Refs.~\cite{ABS,PRL}; the
consequences for the following results are however
rather small.}
$T_c/\Lambda_{\overline{\rm MS}}=1.14$, 
a recent lattice result \cite{TcLa} for the case of pure glue QCD, 
which we take also
for $N_f\not=0$ because lattice data so far showed little sensitivity
of this parameter on $N_f$. In order to have an estimate of the
theoretical uncertainty of our approach we consider a variation
of $\bar\mu$ around the central value by a factor of 2.

Were it not for the trace anomaly \cite{DHLRA} 
$\langle T^\mu_\mu \rangle = {\cal E}-3P \not=0$ 
the pressure would be simply given by $P=(T{\cal S} + \mu{\cal N})/4$
(in the case of massless quarks). 
The correct relation is instead provided by an integration such as
\be
P(T,\mu)=\int_{T_1}^T {\cal S}(T',\mu)\,dT'+P(T_1,\mu)\,.
\ee

\begin{figure}
\epsfysize=4.5cm
\centerline{\epsfbox[70 210 540 520]{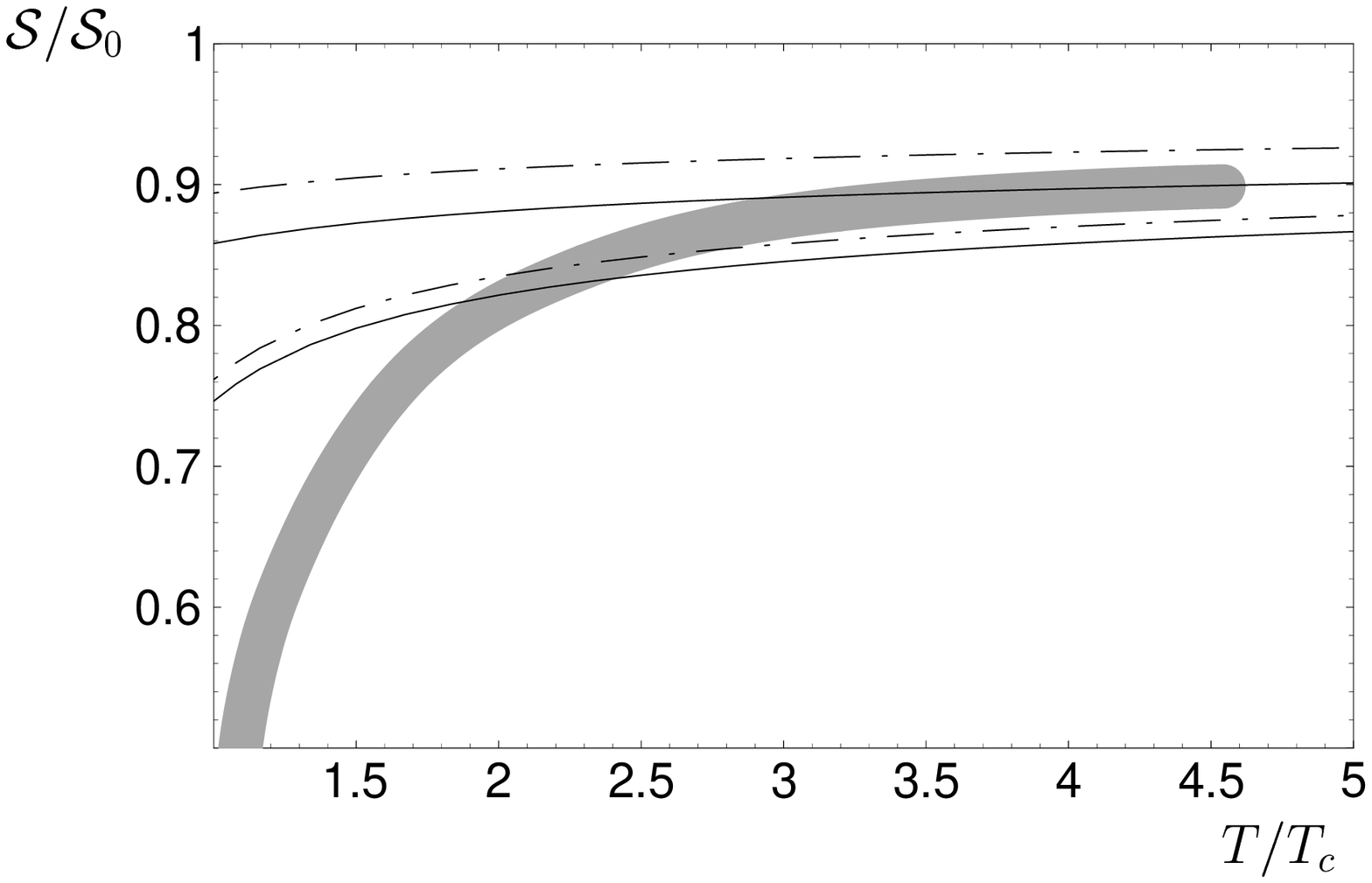}
\epsfysize=4.5cm\epsfbox[5 210 540 520]{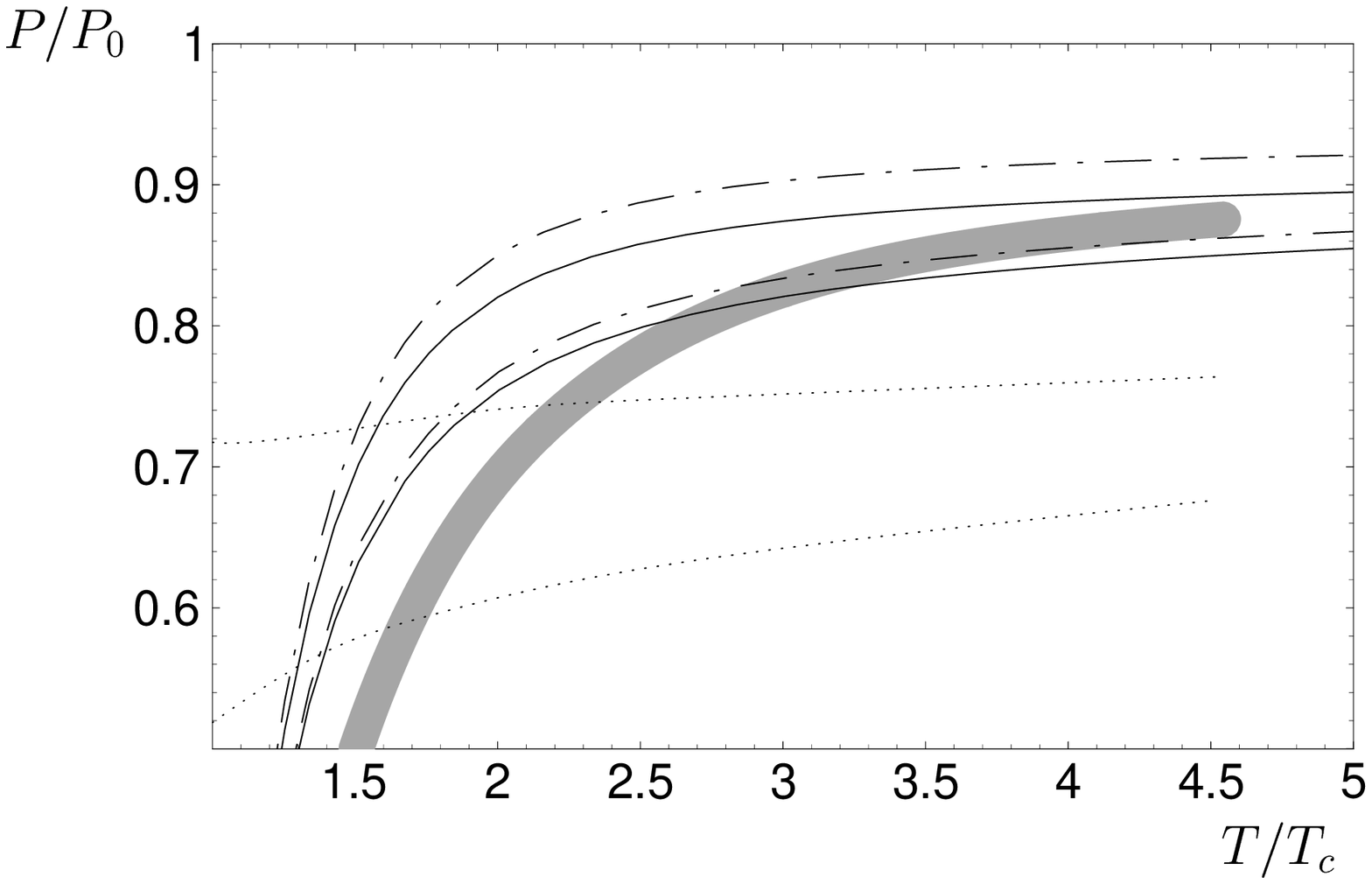}}
\centerline{(a)\hspace{7cm}(b)}
\caption{Comparison of our results (full lines LO;
dash-dotted lines: NLO) for the entropy density (a) and the
pressure (b) of a pure gluon plasma with the lattice results
(grey bands) from Ref.~\cite{Boyd}.
See text for detailed explanations.}
\label{fig1}
\end{figure}

\begin{figure}
\epsfysize=4.5cm
\centerline{\epsfbox[70 210 540 520]{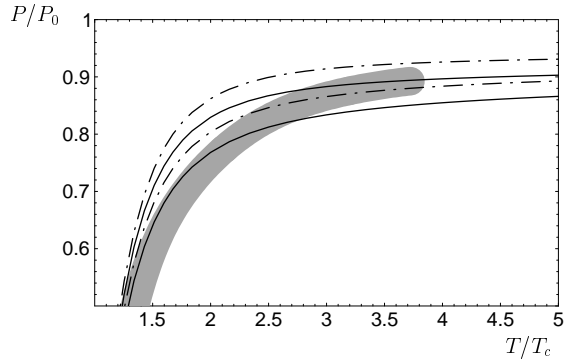}}
\caption{Comparison of our results for the
pressure for a quark-gluon plasma with 
two massless flavours with a continuum extrapolation of
recent lattice results from Ref.~\cite{Karsch}.
}
\end{figure}

At $\mu=0$, the integration constant can be taken from the lattice;
for definiteness we take $T_1=T_c$ and put
$P(T_c,0)=0$. (Nonzero
values as given e.g.\ by the lattice results do not
modify the following results significantly for $T>T_c$.)

In Fig.~1 our results for the entropy (a) and
the pressure (b) of pure-glue QCD are compared with
the lattice results of 
Ref.~\cite{Boyd}\footnote{The more recent
lattice results obtained with a RG-improved lattice action in 
Ref.~\cite{Okamoto} are consistent with the results of Ref.~\cite{Boyd}
within the given errors, but are systematically higher by some
3--4 \%.}.
Full lines give the upper
and lower bounds of the result in the HTL approximation under
a variation of the renormalization scale as described above, the
dash-dotted lines the corresponding one for the NLO approximation,
which include a variation of $c_\Lambda$
in the range $\2$\ldots 2. The lattice results are given
by grey bands with the thickness giving a typical error. 
For comparison, the
result of a direct HTL-resummation of the pressure as reported
in Ref.~\cite{ABS}, which 
fails to incorporate the correct LO term
but reproduces the order $g^3$-part, is given by the dotted line
in Fig.~1b (transcribed to our
value of $T_c/\Lambda_{\overline{\rm MS}}=1.14$).

Fig.~2 gives our results for 
the pressure in the presence of two massless quark
flavours and compares with the (estimated) continuum extrapolation of 
recent lattice results \cite{Karsch}.
The remarkably accurate agreement 
with the lattice result
already  at $T > 2\ldots 3 T_c$
is seen to persist also in the presence of fermions.

In Fig.~3 the result for the trace anomaly is shown for the case of
pure-glue QCD, in comparison with the lattice data from
Ref.~\cite{Boyd}. Here our results are very
weakly dependent on $\bar \mu$ and also LO and NLO results
almost coincide; in fact, at small $T$ our results are dominated
by the integration constant $P(T_c,0)$.
Evidently, the details very close to $T_c$ are not
reproduced, but a substantial part of the lattice result is seen to be
accounted for in our approach. 

Note that
it is not possible to determine a particular $P(T_1)$ from the
requirement that both $p(\a_s)=P(T)/P_0(T)$ and $s(\a_s)=S(T)/S_0(T)$
approach 1 in the limit of $\a_s\to 0$, because the
differential
equation $p(\a_s)+\4\b(\a_s)p'(\a_s)=s(\a_s)$ with
$\b(\a_s)=-\b_0 \a_s^2-\b_1 \a_s^3-\dots$ 
has as homogeneous solutions
$$p(\a_s)_{\rm hom.}=C\e^{-{1\0\a_s}[4\b_0^{-1}+O(\a_s)]},$$ 
and this vanishes at $\a_s=0$
together with all its derivatives! Different $P(T_c,0)$ thus
correspond to exactly the same coefficients of a series expansion
of $p$ in $\a_s$. Because it does not alter the entropy, the physical
interpretation of this additional input is obviously that of
fixing a (strictly nonperturbative) bag constant.

\begin{figure}
\epsfysize=4.5cm
\centerline{\epsfbox[70 210 540 520]{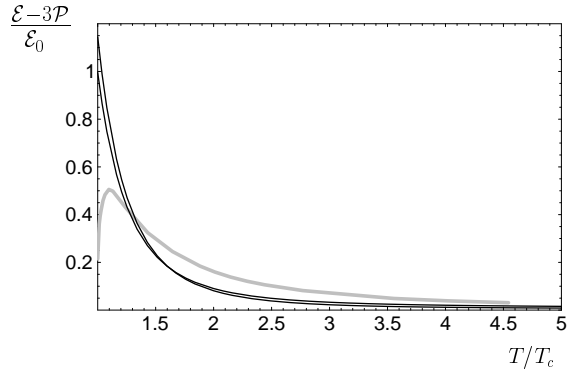}}
\caption{Comparison of our results for the trace anomaly in a
pure gluon plasma with the lattice result from Ref.~\cite{Boyd}.}
\end{figure}

\begin{figure}
\epsfysize=4.5cm
\centerline{\epsfbox[70 210 540 520]{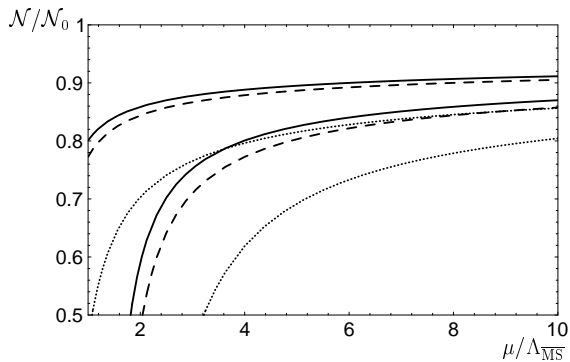}}
\caption{Our result for the baryon density for $N_f=3$ in comparison with
order-$g^2$ (dashed line) and order-$g^4$ (dotted line) perturbation theory.}
\end{figure}

At zero temperature we are lacking the information on the value of
a critical chemical potential
$\mu_c$ and of $P(\mu_c)$. However, we can still calculate
${\cal N}$, and the result of a numerical evaluation for $N=3$,
$N_f=3$ is displayed in Fig.~4. (NLO corrections are not needed now
because the $g^3$-contribution is zero and therefore the HTL result is
trivially correct at this order, too.) Also given are the perturbative
results to order $g^2$ and $g^4$ from Ref.~\cite{Kapusta,FM}
(translated to the $\overline{\rm MS}$-scheme).
A simple quasiparticle model with constant mass $m=M_\infty$
would result in values for ${\cal N}$ only slightly ($\sim 1\%$)
above the perturbative $g^2$-result. Our results are still close
to the latter, but somewhat higher. The perturbative order-$g^4$
results on the other hand are significantly lower with little
overlap with the order-$g^2$ results.
From the good agreement of our above results for the entropy with
lattice results we expect that our approximations for the density
are similarly stable for $\mu$ sufficiently above $\mu_c$.

\section{Outlook}

In our numerical evaluations, we have included NLO effects through
the averaged values (\ref{deltamas}), which suffices to
restore the correct coefficient of $g^3$ contributions. 
A more complete treatment would involve the exact result for
$\delta \Pi_T(\o^2=k^2)$ and $\delta \Sigma_\pm(\o=\pm k)$
in HTL-resummed perturbation theory. Their numerical evaluation
and inclusion in $\cal S$ and $\cal N$ is work in progress.

Another extension of the present results which is under way
is the evaluation
for general $\mu>0$ and $T>0$. There the Maxwell relations
\be\label{Maxrel}
{\6 {\cal S}\0\6\mu}\Big|_T={\6 {\cal N}\0\6T}\Big|_\mu
\ee
are fulfilled up to and including order $g^3$ upon inclusion of the
NLO contributions. At and beyond order $g^4$, (\ref{Maxrel}) constitute
a nontrivial constraint on the renormalization-group flow of
$\a_s(T,\mu)$, determining $P(T,\mu)$ in terms of the initial
data at $\mu=0$, which can be taken from the lattice. This fixes
the equation of state in the entire $\mu$-$T$ plane. A similar
program has been carried out recently in Ref.~\cite{PKSmu} in
simple quasiparticle models with momentum-independent 
transverse-gluon and quark masses. Within our approach, this
can be extended to include both the effects of momentum-dependence
of the thermal masses as well as Landau damping nonperturbatively,
while maintaining equivalence with conventional perturbation theory
up to and including order $g^3$ in the thermodynamic potentials.


\newpage
\begin{thebibliography}{39}
\bibitem{Kapusta}J. I. Kapusta, {\it Finite-temperature field theory}
        (Cambridge University Press, Cambridge, England, 1989).
\bibitem{QCDP}P. Arnold and C. Zhai, Phys. Rev. {\bf D51}, 1906 (1995);
	C. Zhai and B. Kastening, Phys. Rev. {\bf D52}, 7232 (1995);
        E. Braaten and A. Nieto, Phys. Rev. {\bf D53}, 3421 (1996).
\bibitem{Peshier}A. Peshier, B. K\"ampfer, O. P. Pavlenko, and G. Soff,
        Phys. Rev. {\bf D54}, 2399 (1996); A. Peshier, {\tt hep-ph/9809379}.
\bibitem{LH} P. L\'evai and U. Heinz, Phys. Rev. {\bf C57}, 1879 (1998).
\bibitem{KPP} F. Karsch, A. Patk\'os, and P. Petreczky, Phys. Lett. {\bf B401},
        69 (1997).
\bibitem{DHLR2} I. T. Drummond, R. R. Horgan, P. V. Landshoff, and A. Rebhan,
        Nucl. Phys. {\bf B524}, 579 (1998); D. B\"odeker, P. V. Landshoff,
	O. Nachtmann, and A. Rebhan, Nucl. Phys. {\bf B539}, 233 (1999);
	A. Rebhan, {\tt hep-ph/9809215}.
\bibitem{LW} J. M. Luttinger and J. C. Ward, Phys. Rev. {\bf 118}, 1417
        (1960); C. De Dominicis and P.C. Martin, J. Math. Phys. {\bf 5},
	14, 31 (1964).
\bibitem{Baym} G. Baym, Phys. Rev. {\bf 127}, 1391 (1962).
\bibitem{CP2}G. M. Carneiro and C. J. Pethick, 
	Phys. Rev. {\bf B11}, 1106 (1975).
\bibitem{Riedel} E. Riedel, Z. Phys. {\bf 210}, 403 (1968).
\bibitem{BP} J. Frenkel and J. C. Taylor, Nucl. Phys. {\bf B334}, 199 (1990);
E. Braaten and R. D. Pisarski, Nucl. Phys. {\bf B337}, 569 (1990).
\bibitem{MLB}M. Le Bellac, {\it Thermal  Field Theory},
(Cambridge University Press, Cambridge, 1996).
\bibitem{ABS} J. O. Andersen, E. Braaten, and M. Strickland, 
Phys. Rev. Lett. {\bf 83}, 2139 (1999), {\tt 
hep-ph/9905337, hep-ph/9908323}.
\bibitem{PRL} J.-P. Blaizot, E. Iancu and A. Rebhan, Phys. Rev. Lett. {\bf 83},
 2906 (1999).
\bibitem{VB} B. Vanderheyden and G. Baym, J. Stat. Phys. {\bf 93}, 843 (1998).
\bibitem{us} J.-P. Blaizot, E. Iancu and A. Rebhan, to be published.
\bibitem{FM}
B.A. Freedman, and L. McLerran, Phys. Rev. {\bf D16}, 1130, 1147, 1169 (1978).
\bibitem{KKW} O. K. Kalashnikov and V. V. Klimov, Sov. J. Nucl. Phys.
        {\bf 33}, 443 (1981); H. A. Weldon, Phys. Rev. {\bf D26}, 1394 (1982).
\bibitem{P} R. D. Pisarski, Physica A {\bf 158}, 246 (1989);
 	Nucl. Phys. {\bf A498}, 423c (1989);
	J.-P. Blaizot and J.-Y. Ollitrault, Phys. Rev. {\bf D48}, 1390 (1993).
\bibitem{KKR} U. Kraemmer, M. Kreuzer, and A. Rebhan, Ann. Phys. {\bf 201},
        223 (1990) [Appendix]; F. Flechsig and A. K. Rebhan,
	Nucl. Phys. {\bf B464}, 279 (1996).
\bibitem{BR}R. Baier and K. Redlich, {\tt hep-ph/9908372}.
\bibitem{Sch} H. Schulz, Nucl. Phys. {\bf B413}, 353 (1994).
\bibitem{RDeb} A. K. Rebhan, Phys. Rev. {\bf D48}, R3967 (1993).
\bibitem{LP} K. Kajantie {\it et al.}, Phys. Rev. Lett. {\bf 79}, 3130 (1997);
M. Laine and O. Philipsen, {\tt hep-lat/9905004}.
\bibitem{TcLa}O. Kaczmarek, F. Karsch, E. Laermann, and M. L\"utgemeier,
	{\tt hep-lat/9908010} and references therein.
\bibitem{DHLRA} I. T. Drummond, R. R. Horgan, P. V. Landshoff, and A. Rebhan,
        Phys. Lett. {\bf B460}, 197 (1999) and references therein.
\bibitem{Boyd} G. Boyd {\it et al.},
        Nucl. Phys. {\bf B469}, 419 (1996).
\bibitem{Okamoto}M. Okamoto {\it et al.}, {\tt hep-lat/9905005}.
\bibitem{Karsch}F. Karsch, {\tt hep-lat/9909006}
\bibitem{PKSmu}A. Peshier, B. K\"ampfer, and G. Soff,
	{\tt hep-ph/9906305}.

\end{thebibliography}
\end{document}